\documentstyle[12pt]{article}
\newcommand{\be}{\begin{eqnarray}}
\newcommand{\ee}{\end{eqnarray}}

\begin{document}

\vspace{1cm}

\begin{center}

\LARGE{Nucleon and pion electromagnetic form factors in a light-front
constituent quark model}\\

\vspace{1cm}

\large{F. Cardarelli$^{(a)}$, E. Pace$^{(a,b)}$, G. Salm\`{e}$^{(c)}$, S.
Simula$^{(c)}$}\\

\vspace{0.5cm}

\normalsize{$^{(a)}$Istituto Nazionale di Fisica Nucleare, Sezione Tor
Vergata\\ Via della Ricerca Scientifica, I-00133 Roma, Italy \\ $^{(b)}$
Dipartimento di Fisica, Universit\`{a} di Roma " Tor Vergata" \\ Via della
Ricerca Scientifica, I-00133 Roma, Italy \\  $^{(c)}$Istituto Nazionale di
Fisica Nucleare, Sezione Sanit\`{a},\\ Viale Regina Elena 299, I-00161
Roma, Italy}

\end{center}

\vspace{1cm}

\begin{abstract}

Nucleon and pion electromagnetic form factors are evaluated in the
spacelike region within a light-front constituent quark model, where
eigenfunctions of a mass operator, reproducing a large set of hadron
energy levels, are adopted  and quark form factors are considered in the
one-body current. The hadron form factors are sharply affected by the high
momentum tail generated in the wave function by the one-gluon-exchange
interaction. Useful information on the electromagnetic structure of light
constituent quarks can be obtained from the comparison with nucleon and
pion experimental data.

\end{abstract}

\vspace{1cm}

\newpage

\pagestyle{plain}

\indent The measurement of the  electromagnetic (e.m.) form factors of
hadrons represents a valuable tool for investigating in detail their
internal structure. This fact has motivated a great deal of  experimental
and theoretical work, that will increase with the advent of new accelerator
facilities, e.g. CEBAF, yielding unique information on the transition
region from the non perturbative to the perturbative regime of $QCD$
\cite{BL80,IL}. Though the fundamental theory of the strong interaction,
$QCD$, should  be applied for describing hadron structure, the practical
difficulties to be faced  in the nonperturbative regime have motivated the
development of effective theories, e.g. constituent quark (CQ) models, that
in turn  could provide useful hints to model  approximations to the "true"
field theory \cite{KW94}. Aim of this letter is to apply our approach
\cite{pi94,ro95}, based on a relativistic CQ model, to the evaluation of
the  nucleon e.m. form factors in  the spacelike  region, keeping safe the
good description already obtained for the pion form factor. Our model
represents an extension of the  one proposed in  Refs. \cite{CC91,CK95},
where a relativistic treatement of light CQ's is achieved by adopting the
light-front formalism \cite{KP91} and  gaussian wave functions  are assumed
for describing the pointlike CQ's inside the nucleon  (see also
\cite{Dzie}). In particular we have considered: i) hadron wave functions
which are eigenvectors of a light-front mass operator, constructed from the
effective $q\bar{q} $ and $qq$-interaction of Refs. \cite{GI,CI}, that
reproduces a huge amount of energy levels; ii) the configuration mixing,
due to the one-gluon-exchange (OGE) part of the effective interaction,
leading to high momentum components and $SU(6)$ breaking terms in the
hadron wave function; iii) Dirac and Pauli form factors for the CQ's, as
suggested by their quasi-particle nature (cf. \cite{Wei}), summarizing the
underlying degrees of freedom. The comparison of our calculations with the
experimental data on  nucleon and pion form factors will phenomenologically
constrain the e.m. structure of the light CQ's.

\indent  As known (cf. \cite{KP91}), the light-front wave functions of
hadrons are eigenvectors of a mass operator, e.g. ${\cal{M}} = M_0 +
{\cal{V}}$, and of the non-interacting angular momentum operators $j^2$ and
$j_n$, where the vector $\hat{ n} = (0,0,1)$ defines the spin quantization
axis. The operator $M_0$ is the free mass and the interaction term
${\cal{V}}$ is a Poincar\'e invariant. In this letter we briefly present
the formalism for the nucleon only, since the relevant formulae for the
pion can be found in \cite{pi94}. For the three-quark system $M_0^2 =
\sum_{i=1}^3 {q_{i\perp}^2 + m_{i}^2 \over \xi_i}$, where $m_i$ is the CQ
mass, $\xi_i= p_i^+/P^+$ and $\vec{q}_{i\perp}=\vec{p}_{i \perp} - \xi_i
\vec{P}_{\perp}$ are the intrinsic light-front variables. The subscript
$\perp$ indicates the projection perpendicular to the spin quantization
axis and the {\em plus} component of a 4-vector $p \equiv (p^0, {\vec p} )$
is given by $p^+ = p^0 + \hat{ n} \cdot {\vec p}$; finally $\tilde{P}
\equiv (P^+, \vec{P}_{\perp}) = \tilde{p}_1+\tilde{p}_2+\tilde{p}_3$  is
the light-front nucleon momentum and $\tilde{p}_i$ the quark one. In terms
of the longitudinal momentum $q_{in}$, related to the variable $\xi_i$ by
$q_{in} = {1 \over 2}~\left ( \xi_i - { {q_{\perp}^2+m_i^2} \over \xi_i
M_0} \right )$, the free mass  operator acquires a familiar form, viz.
 \be
    M_0 = \sum_{i=1}^3\sqrt{m_i^2 + q_i^2 } =\sum_{i=1}^3~E_i
    \label{5}
 \ee with $\vec{q}_i \equiv ( \vec{q}_{i\perp}, q_{in})$. Disregarding the
color degree of freedom, the light-front nucleon wave function can be
written as follows (see also \cite{CC91,CK95})
 \be
    \langle \{ \xi_i\vec{q_{i\perp}}; \nu_i \tau_i \}|
    \Psi_{N}^{\nu_N}\rangle ~ = ~ \sqrt{{E_1E_2E_3 \over
    M_0\xi_1\xi_2\xi_3}} ~ \sum_{ \{\nu '_i \}} ~ \langle \{ \nu_i
    \}|{\cal{R}}^{\dag}|\{ \nu'_i \} \rangle \langle \{\vec{q_{i}};\nu'_i
    \tau_i \}| \psi_{N}^{\nu_N}\rangle~
    \label{7}
 \ee
where the curly braces $\{ ~~ \}$ mean a list of indices corresponding to
$i=1,2,3$; $\nu'_i (\tau_i)$ is the third component of the quark spin
(isospin); ${\cal{R}}^{\dag}=\left [~\prod_{j=1}^3 R_M^{\dag}
(\vec{q}_{j\perp}, \xi_j, m_j) \right ]$ with  $R_M (\vec{q}_{j\perp},
\xi_j,m_j)$ being the usual generalized Melosh rotation \cite{Mel}.
Disregarding the P and D waves (see below), the equal-time nucleon wave
function $\langle \{\vec{q_{i}};\nu'_i \tau_i\}| \psi_{N}^{\nu_N}\rangle$
is given by
 \be
    \langle \{ \vec{q_{i}};\nu'_i \tau_i\}|  \psi_{N}^{\nu_N}\rangle = ~
    {1 \over \sqrt{2}} ~ \Biggl [ w_S(\vec{p},\vec{k}) ~
    \left(\Phi^{00}_{\nu_N \tau_N}(\{\nu'_i\tau_i \}) + \Phi^{11}_{\nu_N
    \tau_N}(\{ \nu'_i\tau_i \})\right ) ~ + ~ \nonumber \\
    w_{S'}^s(\vec{p},\vec{k}) ~ \left(\Phi^{00}_{\nu_N \tau_N}(\{
    \nu'_i\tau_i \}) - \Phi^{11}_{\nu_N \tau_N}(\{ \nu'_i\tau_i \}) \right
    ) + ~ w_{S'}^a(\vec{p},\vec{k}) ~ \left(\Phi^{01}_{\nu_N \tau_N}(\{
    \nu'_i\tau_i \}) + \Phi^{10}_{\nu_N \tau_N}( \{ \nu'_i\tau_i \} )\right
    ) \Biggr ]
    \label{7.1}
 \ee
where $\vec{p}=\vec{q}_1$ and $\vec{k}=(\vec{q}_2-\vec{q}_3)/2$ are the
Jacobi coordinates for the three-quark system (cf. \cite{CK95}) and the
functions $w_S$ and  $w_{S'}^{s(a)}$ correspond to the $S$ and $S'$
(mixed-symmetry) waves, respectively. Finally, the spin-isospin function
$\Phi^{ST}_{\nu_N \tau_N}(\{ \nu'_i\tau_i \})$ is defined as follows
 \be
    \Phi^{ST}_{\nu_N\tau_N}(\{ \nu'_i\tau_i \}) = \sum_{M_{S}} \langle
    {1 \over 2} \nu'_1 S M_{S}|{1 \over 2} \nu_N \rangle ~
    \langle {1 \over 2} \nu'_2 {1 \over 2} \nu'_3|S M_{S} \rangle ~
    \sum_{M_{T}} \langle {1 \over 2} \tau_1 T M_{T}|{1 \over 2} \tau_N
    \rangle ~ \langle {1 \over 2} \tau_2 {1 \over 2} \tau_3|T M_{T}
    \rangle ~
    \label{8}
 \ee
The normalization of (\ref{7.1}) is $\sum_{\{\nu_i \tau_i \}} \int ~
\prod_{j=1}^3 ~ d\vec{q_j} ~ \delta(\vec{q}_1 + \vec{q}_2 + \vec{q}_3)
\left |\langle \{ \vec{q}_i,\nu_i \tau_i \} | \psi_{N}^{\nu_N} \rangle
\right|^2 ~ = ~1$. The equal-time state $| \psi_{N}^{\nu_N}\rangle$ is an
eigenvector of the transformed mass operator $M = {\cal{R}}
{\cal{MR}}^{\dag} = {\cal{R}}M_0{\cal{R}}^{\dag}+{\cal{R}}
{\cal{VR}}^{\dag}$. Since the free-mass commutes with the Melosh rotation
one has $M = M_0 +V$, where the interaction $V = {\cal{R}}
{\cal{VR}}^{\dag}$ has to fulfil the proper invariance properties, namely:
i) no dependence upon the total momentum and the centre of mass
coordinates, and ii) invariance upon rotations (cf. \cite{KP91}). We have
chosen $M$ equal to the effective hamiltonian proposed by Capstick and
Isgur (CI) \cite{CI}. Thus, our equal-time baryon states $| \psi_{qqq}
\rangle = \sum_{ \{ \nu_i\tau_i \}}(\prod_{j=1}^3 |{1 \over 2} \nu_j
\rangle |{1 \over 2} \tau_j \rangle ) \langle \{ \vec{q_{i}};\nu_i  \tau_i
\}| \psi_{qqq}\rangle$ are eigenvectors of the following mass operator
 \be
    H_{qqq} ~| \psi_{qqq} \rangle ~ \equiv ~ \Bigl [\sum_{i=1}^3
    \sqrt{m_i^2 + q_i^2} + \sum_{i \neq j =1}^3 V_{i j}\Bigr ]  ~ |
    \psi_{qqq} \rangle ~ = ~ M_{q qq} ~ | \psi_{qqq} \rangle
    \label{6}
 \ee
where $M_{q qq}$ is the mass of the baryon, and $V_{ij}$ the CI effective
$q q$  potential, which is composed by a OGE term (dominant at short
separations) and a linear confining term (dominant at large separations).
It should be pointed out that the relativistic mass operator (\ref{6})
reproduces a large set of hadron energy levels \cite{GI,CI} and generates
a huge amount of configuration mixing,  due to the presence of the OGE
part of the interaction. We have evaluated the nucleon form factors using
in Eq.(\ref{7}) the eigenstate of Eq.(\ref{6}), whereas in the current
literature only the effects of the confinement scale have been considered,
through gaussian or power-law wave functions \cite{CC91,CK95,Dzie}. As in
Refs. \cite{GI,CI}, the values $m_u = m_d = 0.220 ~ GeV$ have been
adopted.

\indent The wave equation (\ref{6}) has been  solved by expanding the wave
functions $w_S(\vec{p},\vec{k})$ and $w_{S'}^{s(a)}(\vec{p},\vec{k})$ onto
a (truncated) set of harmonic oscillator (HO) basis states (details will be
given elsewhere \cite{CS}) and  applying to the Hamiltonian $H_{qqq}$ the
Rayleigh-Ritz variational principle. We have checked that the  convergence
for all the quantities considered in this work can be reached including in
the expansion all the basis states up to  $20$ HO quanta. In particular,
the calculated nucleon mass is $M_N=.940~ GeV$ and the percentages of the
various waves are: $p_S=98.1$, $p_{S'}=1.7$ and $p_D=0.2$. The P wave has
been neglected since it does not couple to the main component of the wave
function. As in the case of pseudoscalar \cite{pi94} and vector \cite{ro95}
mesons, the OGE part of the $q q$-interaction, determining the hyperfine
splitting of hadron mass spectra, generates high momentum components in the
baryon wave function \cite{CS,nuc95}.

\indent In what follows we will present the calculations of the nucleon and
pion e.m. form factors performed by using the eigenvectors (\ref{7}) and
the one-body component of e.m. current, which for the nucleon is given by
 \be
    {\cal{I}}^{\nu}_N = \sum_{j=1}^3~I^{\nu}_j=\sum_{j=1}^3 ~ \left ( e_j
    \gamma^{\nu}f^j_1(Q^2) ~ + ~ i \kappa_j {\sigma^{\nu}_{~\mu} q^{\mu}
    \over 2 m_j}f^j_2(Q^2) \right )
    \label{13}
 \ee
where $\sigma^{\nu}_{~\mu} = {i \over 2}[\gamma^{\nu},\gamma_{\mu}]$, $e_j$
is the charge of the j-th quark, $\kappa_j$ the corresponding anomalous
magnetic moment, $f^j_{1(2)}$ its  Dirac (Pauli) form factor, and $Q^2
\equiv - q \cdot q$ the squared four-momentum transfer. In the light-front
formalism, see e.g. Ref. \cite{CC91}, the spacelike e.m. form factors are
related to the plus component of the e.m. current evaluated in a frame
where $q^+=0$; such a choice allows to suppress the contribution of the
pair creation from the vacuum \cite{BL80,FS79}. For the nucleon one has
 \be
    \langle \Psi_{N}^{\nu'_N}| ~ {\cal{I}}^+_N ~ |\Psi_N^{\nu_N} \rangle =
    {\cal{F}}^N_1(Q^2)\delta_{{\nu'}_N\nu_{N}}-i \langle {1 \over
    2}\nu'_{N}| \sigma_2 | {1 \over 2}\nu_{N} \rangle
    \sqrt{\eta_N}{\cal{F}}^N_2(Q^2)
    \label{12}
 \ee
where $\sigma_2$ is a Pauli matrix, ${\cal{F}}^N_{1(2)}$ the Dirac (Pauli)
form factor of the nucleon and $\eta_N=Q^2/(4M^2_N)$. The comparison with
the experimental data will be presented in terms of the Sachs form
factors, i.e. $G_E^N={\cal{F}}^N_1 - \eta_N {\cal{F}}^N_2$ and
$G_M^N={\cal{F}}^N_1 + {\cal{F}}^N_2$. The explicit expression for the
nucleon form factors, including the contributions from $S$ and $S'$ waves,
and for the pion one, including the contribution from the Pauli quark form
factor not considered in \cite{pi94}, will be given elsewhere
\cite{nuc95}. As a check, we have repeated the calculations of the nucleon
form factors of Refs. \cite{CC91,CK95}, where  a simple S-wave gaussian
function was adopted. It should be pointed out that the numerical
calculations involve multifold integrations, carried out through  a Monte
Carlo routine \cite{VEG}.

In order to investigate the sensitivity of the form factor upon the high
momentum tail of the nucleon wave function, we have calculated the nucleon
form factors assuming pointlike quarks (i.e. $f^i_{1} = 1$ and $\kappa_i =
0$) and using different nucleon wave functions $\psi^{(CI)}_N$,
$\psi^{(si)}_N$ and $\psi^{(conf)}_N $. These wave functions are obtained
as solutions of Eq. (\ref{6}) by considering the full interaction
$V_{(CI)}$, the spin-independent part of $V_{(CI)}$ and only its linear
confining part, respectively. The results for the ratio $G_{E}^p/G_D$ (with
$G_D= 1/(1+Q^2/0.71)^2$) are shown in Fig.1. The following comments are in
order: i) the configuration mixing generated by the OGE part of the
interaction sharply affects the nucleon form factor even at low values of
$Q^2$, as in the case of the pion \cite{pi94}, ii) the results of the
calculations performed  with $\psi^{(conf)}_N$ are similar to the ones
obtained with a gaussian wave function, e.g., in Ref. \cite{CK95}. It turns
out that analogous results hold as well for the other e.m. form factors. In
particular, if $\psi^{(conf)}_N$ and $\psi^{(CI)}_N$ are used and
pointlike  CQ's are assumed, the values of the nucleon magnetic moments
result to be $\mu_{p[n]}^{(conf)} =2.74 ~ [-1.60]$ and $\mu_{p[n]}^{(CI)}
=2.28 ~ [-1.19]$, respectively; this means that the configuration mixing
leads to a sizeable underestimation of the nucleon magnetic moments
($\mu_{p[n]}^{exp} = 2.793 ~ [-1.913]$).

Once the high momentum components, generated by the full CI
$qq$-interaction, are taken into account, the nucleon form factors sharply
differ from the dipole prediction (cf. the solid line in Fig.1). A
possible way to solve such a discrepancy is to assume a non trivial e.m.
structure for extended CQ's (see e.g. \cite{Wei,Fos}). Thus, a simple
parametrization of the isoscalar and isovector parts of the CQ form factors
$f^i_{1(2)}$ has been introduced
 \be
    f^{S(V)}_1(Q^2) ={e_u f^u_1 \pm e_d f^d_1 \over (e_u \pm e_d)} =
    {A_1^{S(V)} \over 1+Q^2~B_1^{S(V)}} + {1-A_1^{S(V)} \over
    (1 + Q^2~C_1^{S(V)} / 2)^2} \nonumber \\
    f^{S(V)}_2(Q^2) = {\kappa_u f^u_2 \pm \kappa_d f^d_2 \over (\kappa_u
    \pm \kappa_d)} = {A_2^{S(V)} \over (1+Q^2~B_2^{S(V)} / 2)^2} +
    {1-A_2^{S(V)} \over (1+Q^2~C_2^{S(V)}/3)^3}
    \label{14}
 \ee
Within one-body approximation of the e.m. current the values of $\kappa_u$
and $\kappa_d$ can be fixed by the request of reproducing the experimental
nucleon magnetic moments; in particular, using $\psi^{(CI)}_N$ we have
obtained $\kappa_u = 0.085$ and $\kappa_d = -0.153$.  Differently, the
$12$ constants $A_{1(2)}^{S(V)}$, $B_{1(2)}^{S(V)}$ and $C_{1(2)}^{S(V)}$
have been estimated through a standard minimization procedure, where the
experimental form factors of both nucleon and pion in a wide range of
momentum transfer, reaching for the proton  $Q^2 \approx 30 (GeV/c)^2$,
have been considered. This  wide range can be justified by the
phenomenological attitude adopted in this letter and is also inspired by
the fact that the onset of the perturbative QCD is not definitively
localized \cite{BL80,IL}.

Our calculations (solid lines) are compared with the nucleon and pion data
in Figs. 2 and 3, respectively. It should be pointed out that our results
have been obtained in the framework of the one-body approximation for the
hadron e.m. current, namely by disregarding the two-body currents necessary
for fulfilling both the gauge and rotational invariances (see Ref.
\cite{KP91}). However, Figs. 2 and 3 clearly show that an overall agreement
with the data can be  achieved  by assuming an effective one-body current
that contains CQ's with a structure.  The corresponding CQ form factors,
$f^u_{1(2)}$ and $f^d_{1(2)}$, are presented in Fig. 4 and exhibit a
difference between the u- and d-quark e.m. structure. In order to
illustrate the role played by such a difference, we have also presented the
results (dotted lines in Fig. 2) obtained by assuming $f^S_{1(2)}=
f^V_{1(2)}$, with $ f^V_{1(2)}$ given by the previous  procedure, and the
values of the anomalous magnetic moments fixed as follows:
$\tilde{\kappa}_u / e_u = \tilde{\kappa}_d / e_d = \kappa_V =
\kappa_u-\kappa_d = 0.238$. This prescription, which assumes the same e.m.
structure for u and d quarks, does not change the prediction for the pion
and  helps to illustrate the relevance of the differences between the u and
d quarks for explaining the nucleon data with  accuracy. The nucleon
charge form factors, corresponding to the full calculations (solid lines in
Fig. 2), have the following slopes at $Q^2=0$:
 $dG_{E}^{p(th)}(Q^2)/dQ^2 = -2.83 \pm.20 (c/GeV)^{2}$
[$dG_{E}^{p(exp)}(Q^2)/dQ^2 = -3.0 \pm.09 (c/GeV)^{2}$ \cite{Hoe}] and
 $dG_{E}^{n(th)}(Q^2)/dQ^2 = 0.33 \pm.03 (c/GeV)^{2}$
[$dG_{E}^{n(exp)}(Q^2)/dQ^2 = 0.50 \pm.015 (c/GeV)^{2}$ \cite{Hoe}],
respectively. The theoretical error bars are estimated from the statistical
uncertainties of the Monte Carlo numerical integration procedure.

As to the pion, due to the constraints imposed by the nucleon data, in
particular by $G_E^n$, the overall quality of the fit  is a little bit
lower  ($r_{ch}^{\pi{(th)}} = 0.71 fm$ while  $r_{ch}^{\pi{(exp)}} = 0.660
\pm 0.024~ fm$ \cite{Am}) than the one obtained in Ref. \cite{pi94}, where
only the pion was considered. It should be pointed out that, if the data
for $G_E^n$ are  disregarded in the minimization procedure, an impressive
agreement for all the remaining form factors is obtained, but the predicted
$G_E^n$ becomes quite small \cite{nuc95}.

Finally, it is worth noting that the CQ form factors shown in Fig. 4 yield
a quark radius, defined as $<r_1^{u(d)}>^2=-6~{df_1^{u(d)}(Q^2)/ dQ^2}$ at
$ Q^2=0$, equals to $<r_1^{u(d)}>=0.51 fm~~(0.42 fm)$; such values are
similar to the ones obtained in Refs. \cite{Wei,Pov} and from our
exploratory analysis of the pion data \cite{pi94} (cf. also dotted line in
Fig. 4).

\indent In conclusion, the picture stemming from our analysis points to a
description of the hadron form factors, in a wide range of momentum
transfer, in terms of effective quarks, having a non trivial e.m.
structure. For the calculation of the pion and nucleon form factors we have
adopted the eigenfunctions of a light-front mass operator, reproducing a
large set of hadron energy levels (see Eq.(\ref{6})), and a one-body
approximation for the e.m. current, containing CQ form factors. Within this
framework, the existing pion and nucleon data phenomenologically constrain
the CQ form factors and this fact will allow  parameter-free calculations
of other hadron form factors in the u-d sector. The application of our
approach to the calculation of the magnetic form factor of the $N-\Delta$
transition \cite{nuc95} as well as the evaluation of the corresponding
angular condition \cite{KP91}, yielding an estimate of the effects of
two-body e.m. currents, are in progress.

\vspace{0.5cm}

{\bf Acknowledgement.} We are very indebted to Prof. F. Coester for many
useful discussions, and we gratefully ackowledge  Dr. S. Platchkov for
supplying us with data of the charge form factor of the neutron.

\newpage

\vspace{0.75cm}

\begin{center}

{\bf Figure Captions}

\end{center}

\vspace{0.75cm}

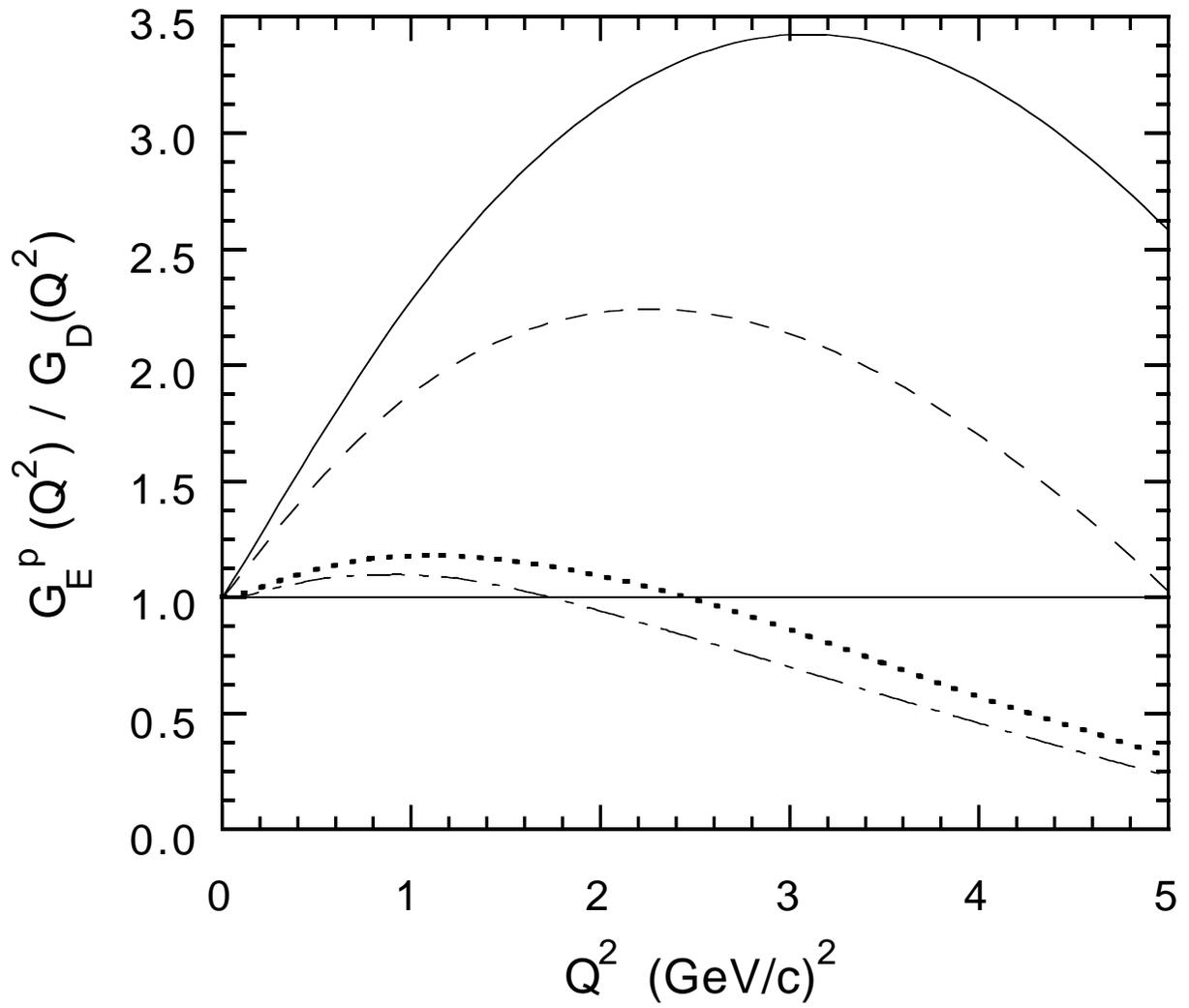
Fig. 1. The proton form factor $G_E^p/G_D$ vs. $Q^2$ for different choices
of the proton wave function, and assuming a pointlike structure for the CQ
(i.e. $f^{u(d)}_1=1$ and $\kappa_{u(d)}=0$). Solid line:  $G_E^p/G_D$
calculated through $\psi^{(CI)}_N$, eigenfunction of Eq.(\ref{6})
corresponding to the full interaction $V_{(CI)}$ of Ref. \cite{CI}; dashed
line: $G_E^p/G_D$ calculated with $\psi^{(si)}_N$, corresponding to the
spin-independent part of the full interaction $V_{(CI)}$; dotted line:
$G_E^p/G_D$  calculated by using $\psi^{(conf)}_N$, corresponding to the
linear confining part of the full interaction $V_{(CI)}$. For comparison
the result of Ref. \cite{CK95} has also been  shown (dot-dashed line).

\vspace{0.75cm}

Fig. 2. - a) The proton form factor $G_E^p/G_D$ vs. $Q^2$. Solid line:
$G_E^p/G_D$ obtained by using: i) the wave function $\psi^{(CI)}_N$
corresponding to the full interaction $V_{(CI)}$ of Ref. \cite{CI}, ii) the
nucleon e.m. current of Eq. (\ref{13}), and iii) the CQ form factors of
Eq.(\ref{14}) (see text). Dotted line: the theoretical $G_E^p/G_D$
calculated assuming $f^S_{1(2)} = f^V_{1(2)}$ and $\tilde{\kappa}_u / e_u =
\tilde{\kappa}_d / e_d = \kappa_V = \kappa_u - \kappa_d = 0.238$ (see
text). Experimental data are from Refs.\cite{Hoe} (full dots), \cite{Wal}
(open squares), \cite{And} (open diamonds)  and \cite{Sil} (full squares).
- b) The same of Fig. 2a, but for $G_M^p/(\mu_p~G_D)$. Experimental data
are from Refs.\cite{Hoe} (full dots), \cite{Wal} (open squares), \cite{And}
(open diamonds)  and \cite{Sil} (full squares).- c) The same of Fig. 2 a,
but for $G_E^n$. Experimental data are from Ref.\cite{Plat} (full dots),
corresponding to the analysis in terms of the N-N Reid soft core
interaction, and from Ref. \cite{Lung} (open squares). - d) The same of
Fig. 2 a, but for $G_M^n/(\mu_n~G_D)$. Experimental data are from
Refs.\cite{Bar} (open diamonds), \cite{Alb} (full diamonds), \cite{Mar}
(full squares), \cite{Lung} (full dots) and \cite{Rock} (open squares).

\vspace{0.75cm}

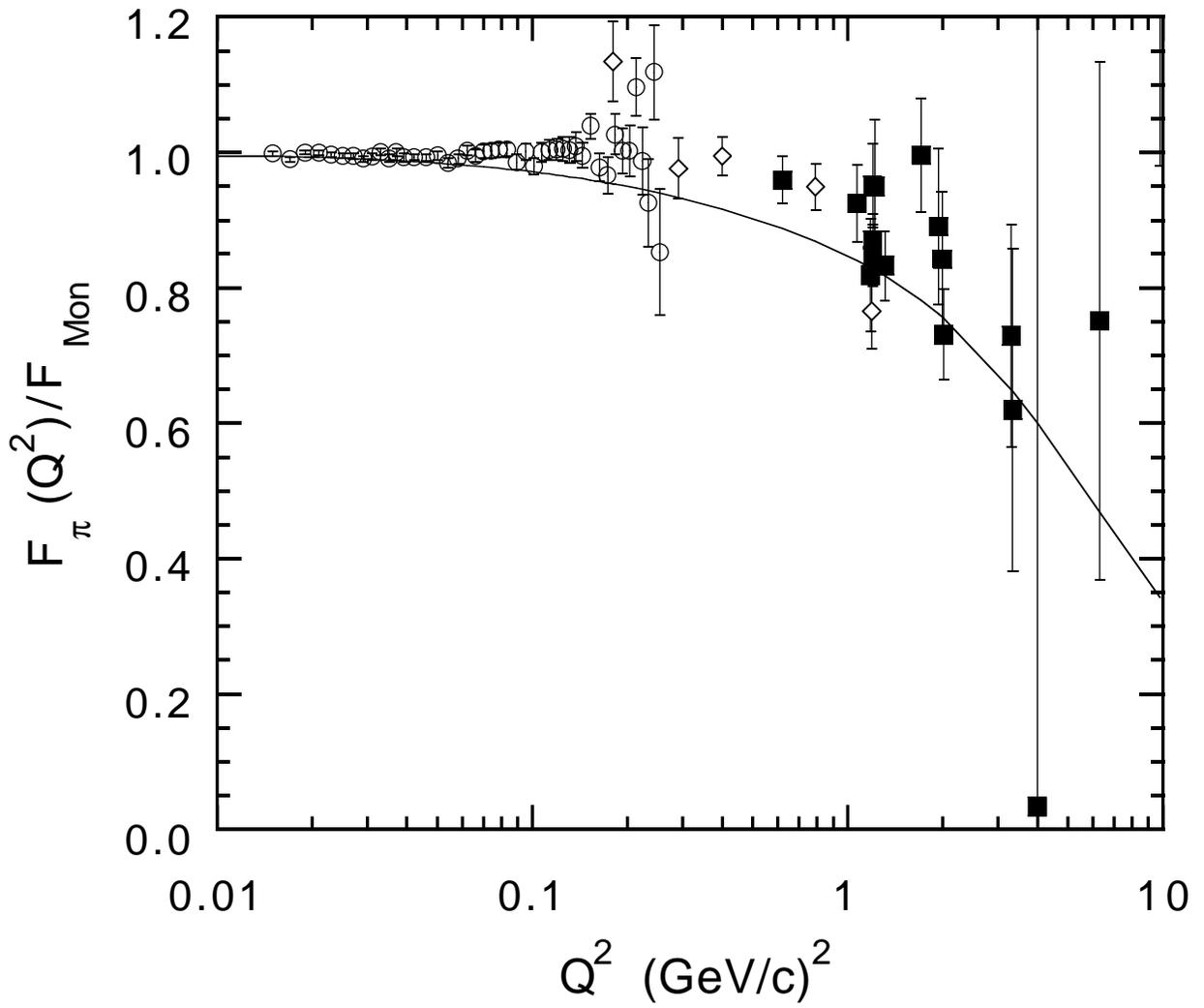
Fig. 3. The charge form factor of the pion $F_{\pi}(Q^2) / F_{Mon}(Q^2)$
vs. $Q^2$, with $F_{Mon}(Q^2) $ $= 1/(1 + Q^2/0.54)$. The theoretical
curve has been obtained by using the wave function of the pion
corresponding to the full interaction of Ref. \cite{GI}, and  the CQ form
factors of Eq.(\ref{14}). Experimental data are from Refs.\cite{Am} (open
dots), \cite{Bro} (open diamonds), and \cite{Beb} (full squares).

\vspace{0.75cm}

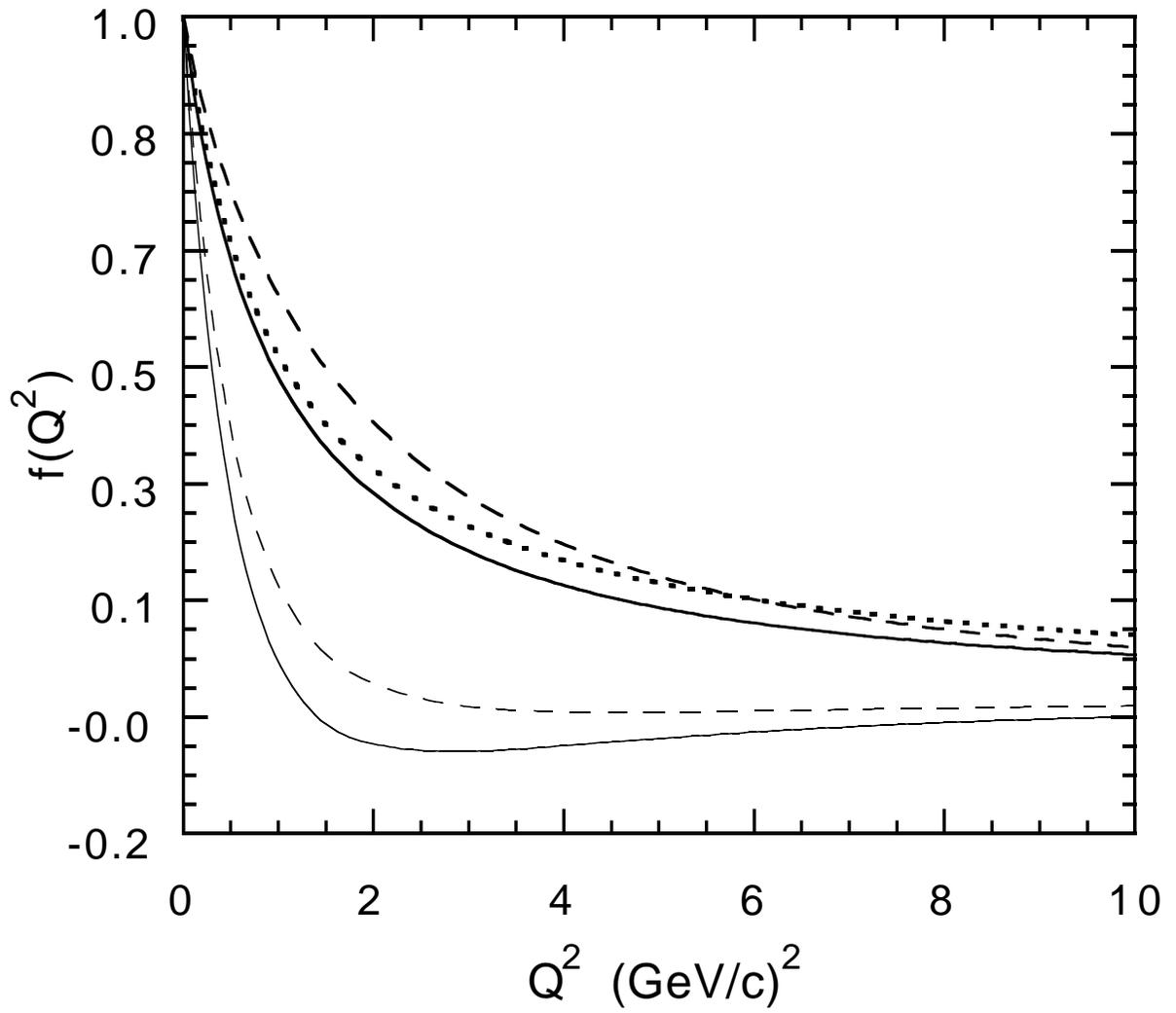
Fig. 4. The constituent quark form factors, extracted from the analysis of
the nucleon and pion form factors (see Figs. 2 and 3), vs. $Q^2$.  Solid
lines represent $f_1^u(Q^2)$ (thick) and $f_2^u(Q^2)$ (thin),
respectively;  dashed lines represent $f_1^d(Q^2)$ (thick) and $f_2^d(Q^2)$
(thin), respectively. For comparison $f_1^u(Q^2)=f_1^d(Q^2)$ obtained in
Ref. \cite{pi94} is also shown ( dotted line).


\vspace{0.5cm}

\begin{thebibliography}{99}

\bibitem{BL80} a) G.P. Lepage and S.J. Brodsky: Phys. Rev. {\bf D22} (1980)
 2157; b) S.J. Brodsky and G.P. Lepage: in {\em Perturbative Quantum
 Chromodynamics}, edited by A.H. Mueller (World Scientific, Singapore,
 1989) p. 93-240.

\bibitem{IL} N. Isgur and C.H. Llewllyn-Smith : Phys. Rev. Lett. {\bf 52}
 (1984) 1080; Phys. Lett. {\bf B 217} (1989) 535; Nucl. Phys. {\bf B 317}
 (1989) 526.

\bibitem{KW94} K.G. Wilson et al.: Phys. Rev. {\bf D 49} (1994) 6720.

\bibitem{pi94} F. Cardarelli, I.L. Grach, I.M. Narodetskii, E. Pace, G.
 Salm\'e and S. Simula: Phys. Lett. {\bf B 332} (1994) 1, and submitted to
 Phys. Rev. D, brief report.

\bibitem{ro95} F. Cardarelli, I.L. Grach, I.M. Narodetskii, G. Salm\'e and
 S. Simula: Phys. Lett. {\bf B 349} (1995) 393.

\bibitem{CC91}  P.L. Chung and F. Coester: Phys. Rev. {\bf D44} (1991)
 229.

\bibitem{CK95} S. Capstick and B. Keister: Phys. Rev. {\bf D51} (1995)
 3598.

\bibitem{KP91} For a review, see B.D. Keister and W.N. Polyzou: Adv. in
 Nucl. Phys. {\bf 20} (1991) 225, and F. Coester: Progress in Part. and
 Nucl. Phys. {\bf 29} (1992) 1.

\bibitem{Dzie} Z. Dziembowski : Phys. Rev. {\bf D37} (1988) 778; I.G.
 Aznaurian: Phys. Lett. {\bf B 316} (1993) 391; H. J. Weber: Phys. Rev.
 {\bf D49} (1994) 3160; F. Schlumpf: Jou. of Phys. {\bf G 20} (1994) 237.

\bibitem{GI} S. Godfrey and N. Isgur: Phys. Rev. {\bf D32} (1985) 185.

\bibitem{CI} S. Capstick and N. Isgur: Phys. Rev. {\bf D34} (1986) 2809.

\bibitem{Wei} U. Vogl, M. Lutz, S. Klimt and W. Weise: Nucl. Phys. {\bf
 A516} (1990) 469.

\bibitem{Mel} H.J. Melosh: Phys. Rev. {\bf D9} (1974) 1095.

\bibitem{CS} F. Cardarelli and S. Simula: to be published.

\bibitem{nuc95} F. Cardarelli, E. Pace, G. Salm\'e and S. Simula: to be
 published.

\bibitem{FS79} L.L. Frankfurt and M.I. Strikman : Nucl. Phys. {\bf B 148}
 (1979) 107; T. Frederico and G.A. Miller: Phys. Rev. {\bf D45} (1992)
 4207; M. Sawicki: Phys. Rev. {\bf D46} (1992) 474.

\bibitem{VEG} G.P. Lepage: J. Comp. Phys. {\bf 27} (1978) 192.

\bibitem{Fos} F. Foster and G. Hughes: Zeit. fur Phys. {\bf C 14} (1982)
 123.

\bibitem{Hoe} G. Hoeler et al.:  Nucl. Phys. {\bf B114} (1976) 505.

\bibitem{Wal} R.C. Walker et al: Phys. Rev. {\bf D49} (1994) 5671.

\bibitem{And} L. Andivahis et al: Phys. Rev. {\bf D50} (1994) 5491.

\bibitem{Sil} A.F. Sill et al.: Phys. Rev. {\bf D48} (1993) 29.

\bibitem{Plat} S. Platchkov et al.: Nucl. Phys. {\bf A510} (1990) 740.

\bibitem{Lung} A. Lung et al: Phys. Rev. Lett. {\bf 70} (1993) 718.

\bibitem{Bar} W. Bartel et al.: Nucl. Phys. {\bf B 58} (1973) 429.

\bibitem{Alb} W. Albrecht et al.: Phys. Lett. {\bf B 26} (1968) 642.

\bibitem{Mar} P. Markowitz et al.: Phys. Rev. {\bf C48} (1993) R5.

\bibitem{Rock} S. Rock et al.: Phys. Rev. Lett. {\bf 49} (1982) 1139.

\bibitem{Am}  S.R. Amendolia et al.: Nucl. Phys. {\bf B 277} (1986) 168;
 Phys. Lett. {\bf B 146} (1984) 116.

\bibitem{Bro} C.N. Brown et al.: Phys. Rev. {\bf D8} (1973) 92.

\bibitem{Beb}  C.J. Bebek et al.: Phys. Rev. {\bf D9} (1974) 1229; Phys.
 Rev. {\bf D13} (1976) 25;  Phys. Rev. {\bf D17} (1978) 1693.

\bibitem{Pov} B. Povh and J. Hufner: Phys. Lett. {\bf B 245} (1990) 653.

\end{thebibliography}
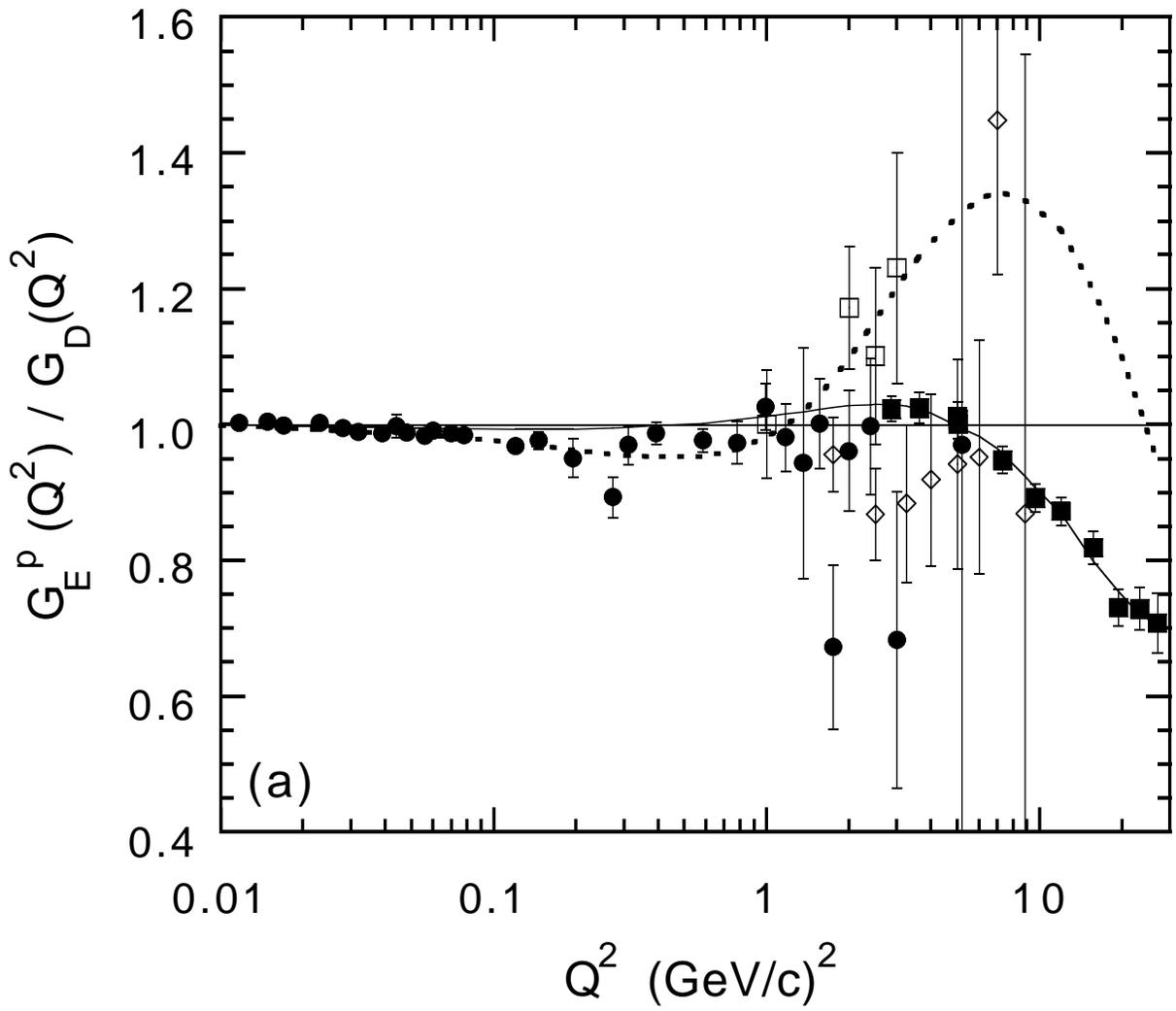
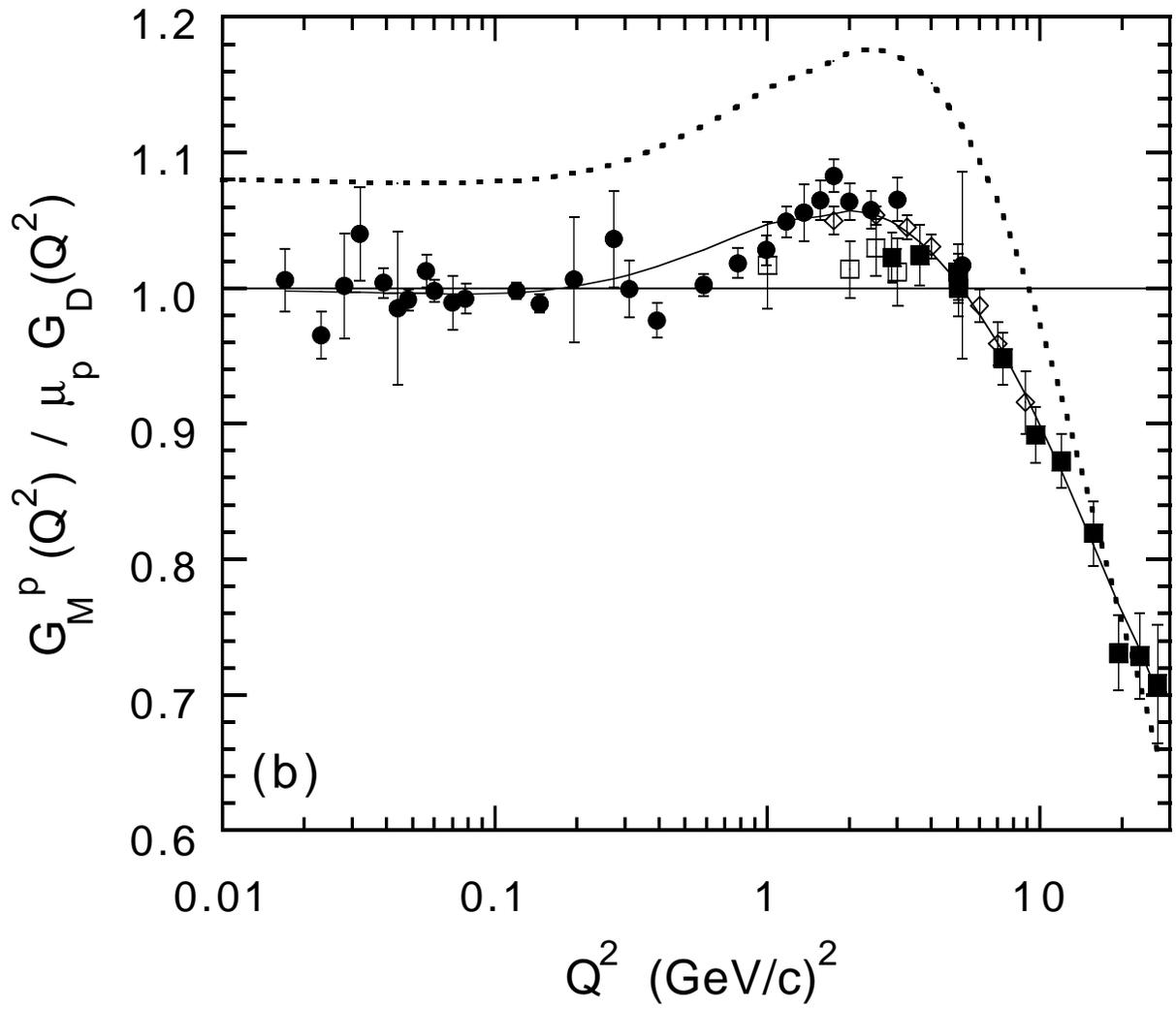
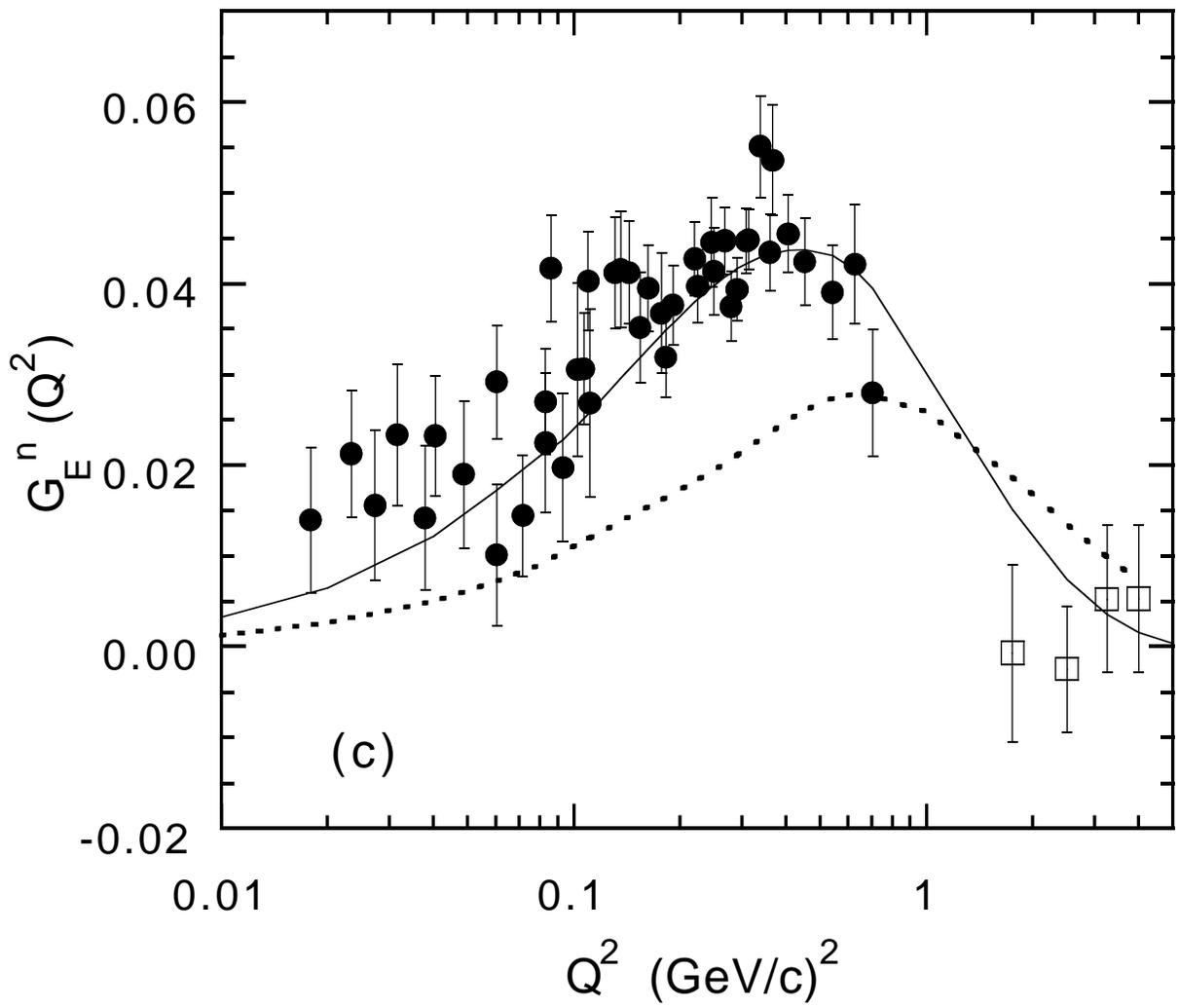
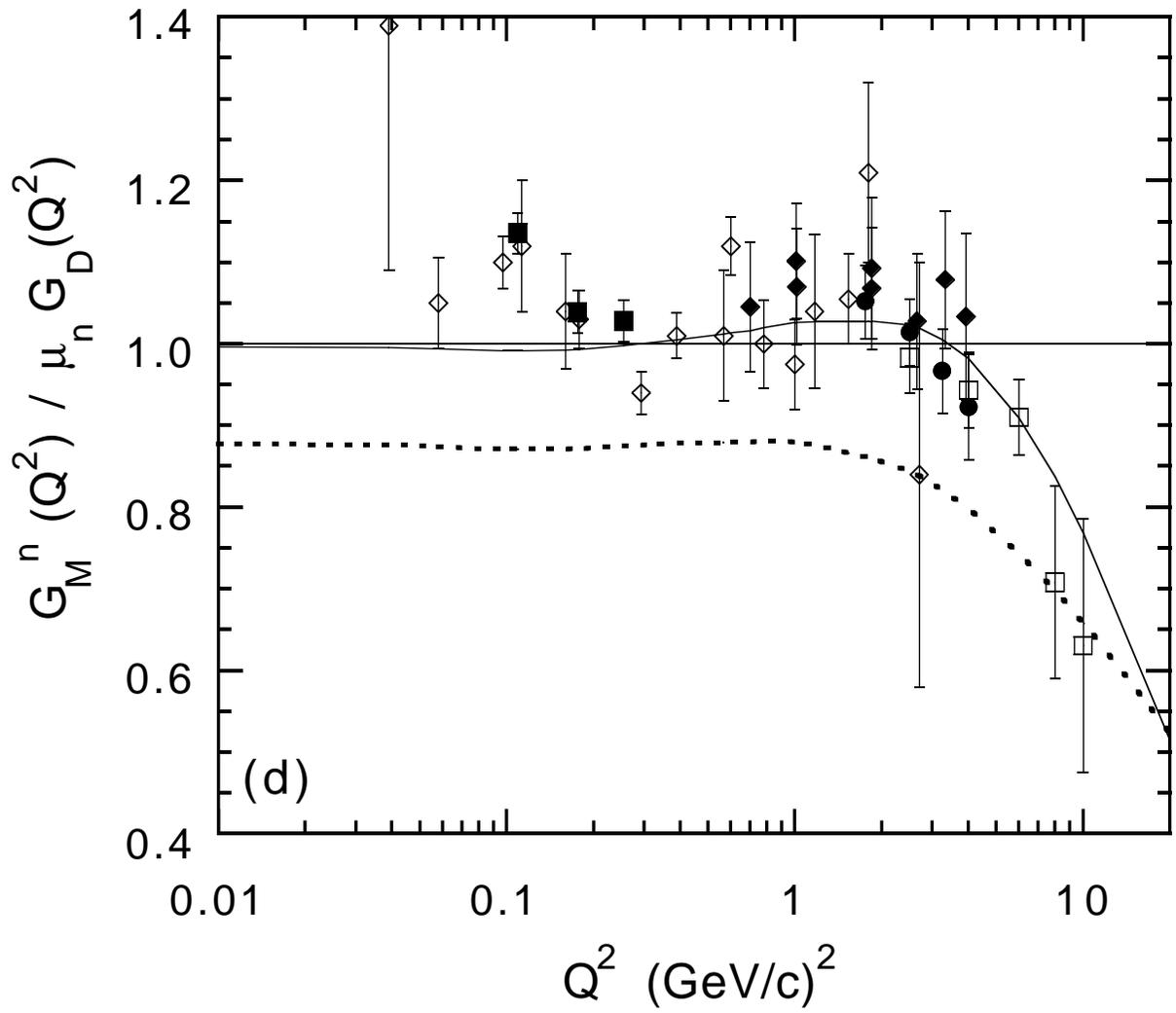
\end{document}